\documentclass[conference]{IEEEtran}
\usepackage{cite}
\usepackage{amsmath,amssymb,amsfonts}
\usepackage{algorithmic}
\usepackage{graphicx}
\usepackage{textcomp}
\usepackage{xcolor}
\def\BibTeX{{\rm B\kern-.05em{\sc i\kern-.025em b}\kern-.08em
    T\kern-.1667em\lower.7ex\hbox{E}\kern-.125emX}}

\usepackage{enumitem}
\newlist{todolist}{itemize}{2}
\setlist[todolist]{label=$\square$}

\begin{document}

\title{Developing a Robust Computable Phenotype Definition Workflow to Describe Health and Disease in Observational Health Research}

\author{\IEEEauthorblockN{1\textsuperscript{st} Jacob S. Zelko}
\IEEEauthorblockA{\textit{HEAT-D} \\
\textit{Georgia Tech Research Institute}\\
Atlanta, Georgia USA \\
jacob.zelko@gtri.gatech.edu}
\and
\IEEEauthorblockN{2\textsuperscript{nd} Sarah Gasman}
\IEEEauthorblockA{\textit{Data for Equity} \\
\textit{Boston Medical Center}\\
Boston, Massachusetts USA\\
sarah.gasman@bmc.org}
\and
\IEEEauthorblockN{3\textsuperscript{rd} Shenita R. Freeman}
\IEEEauthorblockA{\textit{HEAT-D} \\
\textit{Georgia Tech Research Institute}\\
Atlanta, Georgia USA\\
shenita.freeman@gtri.gatech.edu}
\and
\IEEEauthorblockN{4\textsuperscript{th} Dong Yun Lee}
\IEEEauthorblockA{\textit{Biomedical Informatics} \\
\textit{Ajou University School Of Medicine}\\
Suwon, South Korea \\
dongyun90@ajou.ac.kr}
\and
\IEEEauthorblockN{5\textsuperscript{th} Jaan Altosaar}
\IEEEauthorblockA{\textit{University of Tartu} \\
\textit{One Fact Foundation}\\
Tartu, Estonia\\
jaan@onefact.org}
\and
\IEEEauthorblockN{6\textsuperscript{th} Azza Shoabi}
\IEEEauthorblockA{\textit{Janssen Research \& Development} \\
\textit{Johnson \& Johnson}\\
Titusville, New Jersey USA\\
shoaibi.azza@gmail.com}
\and
\IEEEauthorblockN{7\textsuperscript{th} Gowtham Rao}
\IEEEauthorblockA{\textit{Janssen Research \& Development} \\
\textit{Johnson \& Johnson}\\
Titusville, New Jersey USA\\
rao@ohdsi.org}
}

\maketitle

\begin{abstract}
Health informatics can inform decisions that practitioners, patients, policymakers, and researchers need to make about health and disease. Health informatics is built upon patient health data leading to the need to codify patient health information. Such standardization is required to compute population statistics (such as prevalence, incidence, etc.) that are common metrics used in fields such as epidemiology. Reliable decision-making about health and disease rests on our ability to organize, analyze, and assess data repositories that contain patient health data. 

While standards exist to structure and analyze patient data across patient data sources such as health information exchanges, clinical data repositories, and health data marketplaces, analogous best practices for rigorously defining patient populations in health informatics contexts do not exist. Codifying best practices for developing disease definitions could support the effective development of clinical guidelines, inform algorithms used in clinical decision support systems, and additional patient guidelines. 

In this paper, we present a workflow for the development of phenotype definitions. This workflow presents a series of recommendations for defining health and disease. Various examples within this paper are presented to demonstrate this workflow in health informatics contexts.
\end{abstract}


\section{Introduction}

\subsection{Defining Health and Disease Using Phenotypes}

In the context of observational health research, a \textit{phenotype} is a set of observable patient characteristics, such as disease symptoms or diagnoses (e.g., ICD-10 coding, etc.), demographic information (e.g., patient race, sex, income range, and additional social and environmental determinants of health [SEDoH] information), and measurable biomarkers (e.g., blood pressure, HbA1C, etc.). \cite{curcin_why_2020} \cite{burgermaster_psychosocial-behavioral_2022} The documented arrangement of observable patient characteristics that describe a specific patient population is known as a \textit{phenotype definition}. Phenotyping is the process of identifying which patients meet the criteria for a given phenotype definition and is critical for conducting robust and reproducible research. Following a phenotype development process helps to ensure that the results are both generalizable and can be appropriately applied to the intended patient populations. 

Computable phenotype definitions enable the practical application of phenotype definitions. \cite{richesson_framework_2016} Converting phenotype definitions into computable phenotype definitions allow them to be processed by a computer, combining clinically defined criteria and temporal logic in a format consisting of data elements and logical operators. \cite{richesson_electronic_2017} Computable phenotype definitions must accurately capture the clinical presentation of patients and conditions, but are constrained by the availability and quality of data in electronic health records (EHRs) or ancillary data sources (e.g., claims data, disease registries, survey data, etc.). \cite{kostka_chapter_2021}

Although the benefits of establishing robust phenotype definitions are compelling, several concerns must be addressed during the phenotype definition development process. \cite{richesson_framework_2016} \cite{cameron_users_nodate} Researchers have been grappling with such challenges within observational health for over a decade, and have documented several approaches, resources, and best practices in the process. \cite{nichols_construction_2012} The following section discusses some of the central challenges in phenotype development actively being addressed.

\subsection{Challenges in Defining Health and Disease}

\subsubsection{\textbf{Is Patient Data Computable?}}

A challenge commonly encountered while developing a phenotype definition is that phenotypes that would ideally be used to identify patients or diagnose a condition are not consistently present in patient data or accessible in a computable format. While a phenotype definition can clearly define the patient population of interest in several ways (e.g., established coding systems or narrative descriptions), for a phenotype definition to be computable, information must be available in the form of a computable data element. However, when a patient that the phenotype definition intends to capture does not have a previously specified phenotype available in a computable form, complications can arise. The absence of such patient data can occur for many reasons such as a patient did not have contact with a health system, a patient did not report a symptom, a care provider did not order a lab, or care staff failed to record information during a patient encounter. Absence of this data does not necessarily mean that a patient did not have the condition, symptom, or result (absence of evidence is not evidence of absence).

Consider, for example, the racial disparities in the diagnosis and treatment of depression. It is believed that rates of depression are likely similar across racial groups but that patients of color are less likely to be diagnosed with and treated for depression. \cite{blue_cross_blue_shield_racial_nodate} Requiring specific diagnosis codes for inclusion in such a phenotype definition may fail to capture patients of color experiencing symptoms of depression. This results in a phenotype definition that fails to take into consideration demographic disparities, inadvertently contributing to or proliferating inequities in care and obfuscating conclusions about disease presentation, course, outcome, and patients' quality of life. 

\subsubsection{\textbf{Does the Phenotype Capture the Intended Patients?}}

Condition presentation can vary across patient populations, and observational health data systems are prone to significant heterogeneity. Relying solely on HbA1c lab values to identify patients with diabetes, for example, may be precise but could exclude patients taking medication to control blood sugar, those with a history of diabetes that is now in remission, or be irrelevant when applied to claims data, which does not contain exact lab values. A single diagnosis code for diabetes in a patient’s history, however, may not provide enough confidence that this patient should be included. Similarly, consider assumptions of accurate and complete documentation in the case of patient demographic information. \cite{polubriaginof_challenges_2019} Quality of race data can vary significantly; race can be self-reported (considered the gold-standard), recorded based on a clinician's assumption of a patient's race, imputed via algorithm, or in many cases, be left blank. Phenotype definitions that include racial criteria without careful consideration of data source, missingness, and reliability will not be accurate. When developing a phenotype definition, it is important to note how a given condition may present in reality and avoid making assumptions about what data is present and how it is represented in a medical system. \cite{office_of_inspector_general_inaccuracies_2022}

Due to the challenges with consistency of data and patient presentation previously explored, it is unlikely that a single phenotype definition can sufficiently represent even a well-characterized patient population. To capture the intended patient population, phenotype definitions must balance sensitivity and specificity. Using a broad, loosely-constrained phenotype definition may lead to a greater number of patients captured but lower confidence in the resulting cohort, while narrow phenotype definitions may unnecessarily exclude patients. \cite{kostka_chapter_2021} 

Broadly applicable phenotype definitions can be created as a useful starting point for identifying patients in a target population, but it is critical to consider the measurement error associated with a phenotype definition and its impact on the research question, as well as the motivation of the specific research question, before determining which phenotype definition is most appropriate to use. When the patients captured by a phenotype definition for a research study are not equivalent to those clinically intended, findings will suffer from poor internal and external validity \cite{richesson_electronic_2013} For example, a phenotype definition capturing patients with any symptoms of depression could capture patients with Major Depressive Disorder (MDD), as well as patients with Seasonal Affective Disorder (SAD) or Postpartum Depression. Though these conditions may present with similar symptoms, treatments and prognoses are not uniform; failing to address the sensitivity and specificity of the phenotype definition and differentiate appropriately between groups, will result in limited reproducibility due to poor internal and external validity. 

\subsubsection{\textbf{How Specific Should a Phenotype Definition Be?}}

One problem that can arise in developing phenotype definitions is overly constraining phenotype definitions with too many intersecting population axes. \cite{homan2021structural} What is meant by an "axis" in this context, are the various population characteristics by which a general patient population is screened into a final patient cohort (e.g., race, gender, age, etc.). Additionally, intersectionality refers to the intentional overlapping of axes to create a particular patient population. \cite{collins2022black} \cite{crenshaw1991mapping}  Embedded into the notion of intersectional population axes are several concepts such as social context and complexity. \cite{collins2020intersectionality}

An example of this may be when a research group wants to investigate health disparities across diabetic populations in a clinical database. \cite{zelkopilot} A naive approach in defining a phenotype definition to investigate disparity across populations would be to intersect a population by several different axes at once such as by race, gender, and age group. However, if the database only has $10,000$ patients and it represents patients from $5$ different races, $2$ genders, and $10$ age groups, a final patient population may be only a single digit.

This brings to attention a paradox that emerges within observational health research which can be summarized in the question, “How many axes equitably represents a population?” With each added axis to a phenotype definition, correspondingly fewer patients -- and their stories -- will fit into the definition. \cite{homan2021structural} As a result, there exists a tension between creating a phenotype definition that is broad enough to capture a meaningful number of patients but constrained enough to address a specific research objective. Discussions will have to take place within a research team to determine which trade-offs to make given their available data sources.

\subsubsection{\textbf{What Is the Core of a Phenotype Definition?}}

A well-defined research question can guide the development of a phenotype definition; the assumptions at the core of a research question (and subsequent phenotype definition) should be carefully assessed and validated. A research framework that is based on misconceptions about disease presentation or progression; includes inaccurate or unnecessary inclusion or exclusion criteria; or relies too heavily on data that, while 'clean' in research settings, is often 'dirty' in the real-world will result in a phenotype definition that will likely fail to capture the intended patient population.  

Just as misconceptions about disease presentation and progression can delay diagnosis and treatment in clinical settings, these misconceptions can contribute to phenotype definitions that do not account for heterogeneity in patient populations. Consider the variation in symptoms of heart attacks between males and females \cite{douglas_evaluation_1996}. "Typical" symptoms derived from male cohorts, like chest pain, are less precise in identifying angina, obstructive coronary artery disease (CAD), and acute myocardial infarction in females \cite{keteepe-arachi_cardiovascular_2017}. If the core of a phenotype definition is comprised solely of typical symptoms, a cohort identified using this definition would be biased towards typically-presenting patients. Evaluating positive predictive value and negative predictive value is essential in all cases, especially when assumptions may not match reality.

Similarly, assumptions about the correlative and causal relationships of health data elements can result in phenotype definitions that are biased, incomplete, or incorrect. Consider potential phenotype definitions for identifying patients with eating disorders. Eating disorders are often perceived by clinicians as highly correlated to race and only affecting white women \cite{sala_race_2013}. In reality, prevalence across racial and ethnic groups is likely more similar than different. \cite{rodgers_eating_2018} Additionally, diagnostic criteria for eating disorders has historically been focused on weight and weight loss rather than disordered behaviors, with evidence-based treatment frequently withheld by clinicians and insurers if relatively arbitrary weight-based criteria are not met \cite{thomas_eighty-five_2009}. A phenotype definition that assumes correlation (as in the assumed correlation between low weight and disordered eating behaviors), or one that imposes weight restrictions may be appropriate for some research scenarios, but may fail to capture the intended population.

\section{Previous Work}

There is consensus in the literature that standardization of phenotype definition development “can enable the comparison of results across studies, and ensure that all patients can be reliably identified and offered evidence-based treatment options and opportunities for research”. \cite{richesson_framework_2016} Prescriptions for what this standardization entails vary, but priorities include: 

\begin{itemize}
    \item Conform to existing standards and vocabularies (such as SNOMED, LOINC, etc.) \cite{mo_desiderata_2015} 
    \item Create explicit, reproducible, reliable, and valid phenotype definitions for the intended use \cite{cameron_users_nodate}
    \item Be thoroughly documented and transparently applied \cite{richesson_framework_2016}
\end{itemize}

Efforts have been made to streamline the process of phenotype definition development, maintenance, evaluation, and reuse. Initiatives such as the Observational Health Data Sciences and Informatics (OHDSI) open science collaborative \cite{noauthor_ohdsi_nodate}, Patient-Centered Clinical Research Network (PCORNET) \cite{noauthor_pcornet_nodate}, Health Data Research UK (HDRUK) \cite{noauthor_phenotype_nodate}, and Phenotype Knowledgebase (PheKB) \cite{noauthor_phenotypes_nodate} all have developed guidelines around phenotype definition creation and management. While these standards address submission format, phenotype definition components, and methods for assessing and documenting phenotype definition validity, they are often focused on communicating phenotype definitions in computable formats; efforts related to the full process of creating a phenotype definition from an initial research question to final patient population needed for analyses are in process, but have not yet fully materialized.

\section{Considerations and Recommendations}

\begin{figure*}
\centerline{\includegraphics[width=\textwidth]{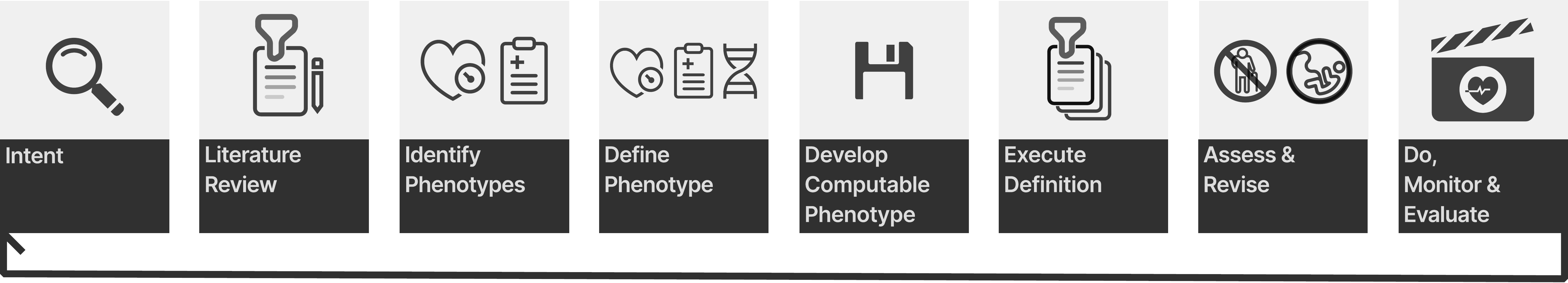}}
\caption{The Phenotype Workflow enables reproducible, replicable biomedical research at scale to answer research questions across domains including epidemiology, biomedical informatics, public health, and others. The steps involved in this continuous process are as follows: 1) define the core motivating factor for the research question, idea, or defined need; 2) review of journals or publications related to clinical informatics, medical anthropology, or public health to find relevant phenotypes; 3) identify relevant phenotypes for use in; 4) translate the needs in creating a phenotype definition by reviewing a disease or condition and considering various constraints for a population; 5) determine if a vocabulary is mapped sufficiently, a data dictionary is identified, and assess feasibility of computable phenotype development; 6) execute the phenotype definition using a query, prediction algorithm, or other method; 7) assess and revise the phenotype definition based on its performance in accurately identifying patients for a phenotype definition; 8) Leverage the results of the computable phenotype definition to drive policy decisions, build clinical insights, and further improve the existing phenotype definition through ongoing monitoring and evaluation. A computable phenotype definition is never fully complete and as a result, the end of this diagram loops back to the start to restart the process to improve phenotype definition as clinical understanding grows.}
\label{general_workflow}
\end{figure*}

Phenotype definitions can enable the generation of high-quality and reliable evidence from observational health data. Establishing common principles, as shown in \ref{general_workflow}, for phenotype definitions helps mitigate challenges detailed earlier in this paper, and creates consistency in design and quality. Based on the activities of the OHDSI Phenotype Development Workgroup and review of the current literature, the following considerations and recommendations were drafted to support phenotype definition development (note: these recommendations are not in perfect alignment with current practices within the OHDSI ecosystem but serve as inspiration for these considerations in the workflow).

\subsection{Identifying Phenotypes for a Phenotype Definition}

To begin constructing a phenotype definition, one must establish a motivation for developing a phenotype definition. Deciding on a motivation, whether in the form of a research question or general goal, can impose useful constraints in the definition's construction. This can assist in identifying phenotypes, including symptoms, test values, and diagnosis codes. Selected phenotypes can be reviewed or validated by a literature review, and will inform the definition by providing options for identifying the intended patient population. For example, if chronic kidney disease is defined by the level of eGFR in claims data, patients who do not have a lab value but are being treated for chronic kidney disease would be excluded. If chronic kidney disease is defined based on its associated symptoms, such as edema and poor appetite, other patients would be included, reducing specificity but increasing sensitivity. Therefore, it is necessary to establish both the intent behind the phenotype definition and the core concepts that can be evaluated for its clinical accuracy and ability to adequately capture intended patients. 

\subsection{Constraining the Phenotype Definition}

Next, one can constrain a phenotype definition by a disease's typical duration, progression and prognosis to capture particular patients of interest. For a disease's duration, specific phenotypes may become more or less relevant over time. Understanding how expressions of a phenotype progress over time, especially those that are included in a phenotype definition, is important for specifying temporal restrictions. Prognosis, or likely outcome, can help determine when patients no longer meet the criteria of a phenotype definition. Using these disease-specific constraints can add much needed specificity throughout the phenotype definition development process.

Furthermore, other constraints can be added, such as disqualifiers and strengtheners, which operate more heuristically based on a disease of interest. \cite{ohdsi2019book} Disqualifiers are defined as mutually exclusive conditions that would prevent a person from having the phenotype of interest (such as appendicitis occurring after an appendectomy). Including disqualifiers in a phenotype definition can strengthen the confidence in the patients that remain. Strengtheners are defined as indicators that increase the likelihood that a person has the defined clinical concept (e.g., patient has a prescription for insulin and no diagnosis, in the data, for diabetes). By including such criteria in the phenotype definition, assessing what proportion of initially captured patients are impacted by strengtheners and disqualifiers can provide critical feedback regarding the validity of the phenotype definition.

\subsection{Defining Logic for Retrieving Patient Characteristics in a Phenotype Definition}

After phenotypes for a phenotype definition have been identified, the next step is to create its logic description which overlaps slightly with previously mentioned constraints.  This step consists of defining entry events, index dates, inclusion or exclusion criterion, and exit criterion. The entry event can be any event recorded that adds a patient into a patient cohort such as a condition diagnosis, performed procedures, prescribed medication, lab measurement, or visitation. The index date could be different from the entry event in that it is the time point at which patients are eligible to enter a cohort based on a phenotype definition. Inclusion and exclusion criteria further restrict the patient cohort and encompass strengtheners and disqualifiers. Finally, exit criteria is when a patient no longer qualifies to be a member of a cohort defined by a phenotype definition. \cite{ohdsi2019book} 

For example, suppose we have a phenotype definition for hypertension and we would like to identify patients with hypertension who are starting anti-hypertensive medications for the first time. A first time exposure to anti-hypertensive medications would be the entry event, The date of first exposure to anti-hypertensive medications is the index date. A requirement for the diagnosis of hypertension before the index date is the inclusion criteria. Exposure to anti-hypertensive medications before the index date is the exclusion criteria. Finishing exposure to anti-hypertensive drugs is the exit criteria. In this way, we have built into the original hypertension phenotype definition, logic that more clearly defines what patients are precisely being considered. 

\subsection{Converting Phenotype Definition to a Computable Form}

Combining the relevant phenotypes with a logic description results in a phenotype definition. It is then necessary to verify that this phenotype definition can be implemented as a computable phenotype definition. Core phenotypes and concepts included in a phenotype definition must be consistently present in the data, and challenges discussed in previous sections (assumptions about disease presentation and progression, phenotype definition sensitivity and specificity, etc.) should be considered. Finally, a phenotype definition can be implemented in a computable form (whether in the form of software or database querying tools) and executed against a database.

\subsection{Checklist for Phenotype Development Workflow}

To summarize this process, the following checklist can be used in developing a phenotype definition:

\begin{todolist}
    \item Intent
    \begin{todolist}
    \item Decide on a motivation to guide phenotype definition construction
    \end{todolist}
    \item Literature Review
    \begin{todolist}
        \item Identify phenotypes
        \item Review or validate selected phenotypes 
    \end{todolist}
    \item Identify Phenotypes
    \begin{todolist}
        \item Specify phenotypes
        \item Add constraints
    \end{todolist}
    \item Translate needs to phenotype definition
    \begin{todolist}
        \item Define disqualifiers and/or strengtheners 
        \item Combine relevant phenotypes with a logic description
    \end{todolist}
    \item Develop Computable Phenotype
    \begin{todolist}
        \item Review what concepts are consistently present in data
        \item Implement phenotype definition in a computable form
    \end{todolist}
    \item Execute Definition
    \begin{todolist}
        \item Execute computable phenotype definition on database
    \end{todolist}
    \item Assess \& Revise
    \begin{todolist}
        \item Consider assumptions within phenotype definition
        \item Revise phenotype definition as needed
    \end{todolist}
    \item Do, Monitor, \& Evaluate
    \begin{todolist}
        \item Answer research question or achieve goal.
        \item Repeat phenotype definition checklist as appropriate
    \end{todolist}
\end{todolist}

\section{Discussion}

Phenotype definitions have the potential for robust application in observational health research, healthcare systems, and public health. However, this must be done with intentionality to avoid inappropriate patient cohorts that skew research, mask health disparities, and exacerbate health inequalities. Systematic phenotype definition development, as proposed and discussed in this paper, is the starting point for creating an accurate and appropriate phenotype definition that answers sound research questions. The development process, like any scientific research, should have reproducibility goals for both the methodology and result.

As shown in the checklist, phenotype definition development starts with a motivating factor and a review of pertinent literature by an interdicsiplinary group. Computable phenotype definitions are to be carefully developed with regard to the veracity and variability of the available data. Soliciting input from individuals - such as clinicians, informaticists, medical anthropologists, health policy experts and others - helps ensure definition applicability and accurate clinical representation. In the case of preeclampsia, a condition with a highly variable clinical concept, the primary inclusion criteria could be a diagnosis of pregnancy at or beyond 20 weeks followed by positive findings for hypertension, pulmonary edema, and/or proteinuria. Any one of these conditions alone may constitute a different phenotype definition but together, in a specific order, they make up a logical definition of preeclampsia which is associated with increased risk of severe maternal morbidity and infants born small for their gestational age. \cite{gregg_preeclampsia_2014} \cite{mayo_clinic_staff_preeclampsia_2022} \cite{mendez-escobarelena_health_2022} After an initial definition has been outlined, inclusion and exclusion criteria are applied to balance sensitivity and specificity as appropriate. This critical step of delineating true positives and true negatives has real-world implications for managing costs of care, preventing over and under treatment, and maintaining or improving a patient’s quality of life.

After initial development and execution, computable phenotype definitions should undergo intermittent review, denoted by the cyclic nature of \ref{general_workflow}. This includes monitoring and evaluation to confirm performance and positive predictive value (PPV) over time. Computable phenotype definitions should not be seen as static due to evidence around key disease processes, characteristics, and associations often continues to emerge.

\section{Conclusion}

In summary, paper has presented several considerations and recommendations in a foundational framework for phenotype definition development that is aimed at formalizing phenotype definition development for equitable use across a variety of settings. Instead of being an exhaustive framework, the intention is to outline a process that prompts phenotype developers to consider significant concepts such as data quality and availability, timing of inclusion and exclusion criteria, validity, and reproducibility. The goal is to prevent both common and critical mistakes that might negatively impact patients and the broader body of knowledge across interdisciplinary research groups.

There are several potential next steps for building upon the workflow proposed in this paper. Refinement for the proposed workflow based on feedback from the greater health informatics research community can help further improve the workflows utility and effectiveness. Future studies or investigations based on this workflow could be done to scrutinize how well the workflow can be used in practice. In conclusion, the steps described within this workflow can provide the foundations to concretely define disease and bring justice to diverse patient populations with phenotype definitions that truly represent them. 

\section*{Acknowledgments}

This work is dedicated to the Observational Health Data Sciences and Informatics (OHDSI) open science community for their leadership in the space of observational health research and their collective support. Particularly, we would like to thank and recognize the OHDSI Phenotype Development Workgroup leadership team, Gowtham A. Rao, Azza A. Shoabi, and Patrick Ryan, for their support; the original conceptualization of the suggested guidance for phenotype development presented within this paper; and their feedback and commentary while finalizing this paper. Clapper clipart provided by DinosoftLab from The Noun Project; all other icons provided by the Health Icons project.

\bibliography{IEEEabrv,references}

\end{document}